# DYNAMICAL SCREENING FUNCTION AND PLASMONS IN THE WIDE HgTe QUANTUM WELLS AT HIGH TEMPERATURES

Evgen O.Melezhik[1*], Fiodor F. Sizov[1], Nikolai N. Mikhailov[2], Joanna V.Gumenjuk-Sichevska[1]

[1]V.E. Lashkaryov Institute of Semiconductor Physics NAS of Ukraine, Kyiv, Ukraine

[2] Institute of Semiconductor Physics SB RAS, Novosibirsk 6300090, Russia

* emelezhik@gmail.com

*Abstract*—Dynamical screening function of the two-dimensional electron gas in wide HgTe quantum well (QW) has been numerically modelled in this work. Calculations were provided in the Random Phase Approximation (RPA) framework and were based on Lindhard equation. Our simulations directly incorporated non-parabolicity of bulk 2D carriers' spectrum, which was obtained by full 8-band k.p method. In the literature exists data that transport properties of HgTe QWs are explained by graphene-like screening. We provide the comparison of the screening function for the Schrodinger fermions in the inverted bands HgTe QW with the appropriate screening function for graphene monolayer with the Dirac fermions. In addition, the dependencies of HgTe-specific screening function on temperature, scattering wave-vector and frequency are studied with the purpose to study the transport properties under high frequency radiation the QWs structures to be used as THz detectors. Plasmon frequencies of 2DEG in HgTe quantum well under study were calculated in the long-wavelength limit for T = 77 K.

I. INTRODUCTION

The development of novel detectors operating in the terahertz (THz) spectral range is one of the important challenges of modern optoelectronics.

HgTe quantum wells with the width larger than ~6.7 nm, have inverted bands order inside the well and therefore semimetal conductivity type seems attractive for THz signal detection because the QWs can be characterized by high electron mobility and high electron concentrations even at liquid nitrogen temperatures [5]. Electron mobility order of $10^5$ cm$^2$/Vs was experimentally reached in such structures at T = 77 K[14, 15], while theoretical estimations predict that such mobility can be increased up to 6-8*$10^5$

cm$^2$/Vs. Moreover, they have much lower resistivity and lower thermal noise in comparison with the direct bands order HgCdTe QWs [1]. All these features of HgTe quantum wells (QWs) with inverted bands order make them so important for various applications in the field of THz detection at the moderate cooling regime. In this article, we deal with the HgTe quantum wells with inverted bands order at the temperature 77 K at which such structures can be exploiting as THz detectors.

The phenomena of Coulomb charge screening of carrier scattering by two-dimensional electron gas (2DEG) has the key importance for theoretical study of high frequency transport and radiation absorption in materials with the high electron concentration. The form of the dielectric screening function sufficiently relies on the energy dispersion law for two-dimensional carriers. Due to the strong nonparabolicity of such dispersion law in HgTe QWs, the dielectric function for two-dimension systems with parabolic dispersion cannot be applied to our case. There exists experimental evidence that the dielectric screening function for HgTe QWs with the stabilized surface state transport in topological insulator phase is close to the dielectric function of graphene with the linear energy dispersion law [2]. In contrast to the case of Dirac fermions, there does not exist any analytical form for dielectric screening function of 2DEG in HgTe inverted bands QWs for the Schrodinger fermions. This case occurs when the temperature is high enough to cancel edge states transport, or doping level is far from the intrinsic case, and electron states are heavy degenerated. This case also differs from parabolic dispersion case due to a strong band mixing and high density of states. However, to create detectors operating with moderate cooling, this case is topical. Thus, knowledge of dielectric screening function for the inverted bands HgTe QW with its unique dispersion law is of key importance for further modeling of any transport processes in it.

In the literature there exist calculations of dynamical screening function for HgTe quantum wells near the phase transition [12]. However, this model based on the four-band calculations of energy spectrum, which are applicable only at the liquid helium temperatures, and cannot be used even at 77 K.

Current work dedicates to the numerical modeling of dielectric screening function and plasmon frequencies for HgTe QW in Random Phase Approximation (RPA) formalism, which was previously successfully applied to the modeling of such function in graphene [3]. In the RPA, electrons respond only to the total electric potential that is consisting of an external perturbing potential and a screening potential. In this assumption, the perturbing potential oscillates at a frequency of external radiation source *w*, so that the model gives self-consistent dynamic dielectric function, which is called the Lindhard dielectric function.

In this work all simulations were provided for $Hg_{0.15}Cd_{0.85}Te$ / HgTe / $Hg_{0.15}Cd_{0.85}Te$ quantum well of 20 nm width, at the moderate cooling temperature, for arbitrary values of scattering wave-vector *q* and frequency *w*. Quantum well had two n-type delta-doped layers in the barriers, at the distance of 10 nm from QW interfaces. Width of delta-doped layers was 15 nm, concentration of dopant was $10^{17}$ cm$^{-3}$. We assume that all free electrons from the delta-doped layers are injected into the well.

## II. Dielectric screening function in RPA formalism

### A. *Lindhard equation*

Numerical simulation of HgTe QW dielectric screeningfunction is based on the well-known Lindhard equation for dynamical dielectric function:

$$\epsilon(\vec{q},w) = 1 - V_q \sum_{\vec{k}} \frac{f_{\vec{k}-\vec{q}} - f_{\vec{k}}}{\hbar(w+is) + E_{\vec{k}-\vec{q}} - E_{\vec{k}}} \quad (1)$$

Here $f_k$ – Fermi-Dirac distribution function, $E_k$ is the energy dispersion law for the given system, *s* is a positive infinitesimal, that is taken in the limit of going to +0 to produce the real. $V_q$ is the Fourier transform of the Coulomb energy of 2DEG in a vacuum. The expression for $V_q$ is [3]:

$$V_q = (2\pi e^2/\kappa q) F(q) \quad (2),$$

where $\kappa$ is the dielectric constant of the lattice. The form-factor *F(q)* will be taken equal to *1*, which corresponds to the strict two-dimensional limit [4].

Suggesting the case, when only ground level of quantum well is populated, and for two-dimensional medium, the last equation can be rewritten as:

$$\epsilon(\vec{q}, w) = 1 - \frac{e^2}{\pi \kappa q} \iint d^2\vec{k} \frac{f_{\vec{k}-\vec{q}} - f_{\vec{k}}}{\hbar(w+is) + E_{\vec{k}-\vec{q}} - E_{\vec{k}}} \qquad (3),$$

where we have substituted the value for $V_q$ and converted the sum over all occupied electron states to the integral over 2D electron wave-vector $\vec{k}$.

Using the Dirac identity:

$$\lim_{s \to 0} \frac{1}{X+is} = p.v. \frac{1}{X} - i\pi\delta(X) \qquad (4),$$

where p.v. denotes the principal value of an integral and δ(X) is the Dirac delta function, we can obtain the real and imaginary parts of the dielectric function.

*B. Real and imaginary parts of dielectric function*

Using the Polar coordinates and neglecting the dependence of electron energy on the orientation of its wave-vector, we can write real and imaginary parts of dielectric function as:

$$\epsilon_{RE}(\vec{q}, w) = 1 - \frac{e^2}{\pi \kappa q} \iint dk d\theta \cdot k \frac{f_{|\vec{k}-\vec{q}|} - f_{\vec{k}}}{\hbar w + E_{|\vec{k}-\vec{q}|} - E_{\vec{k}}}$$

$$\epsilon_{IM}(\vec{q}, w) = -2 \frac{e^2}{\kappa q} \int dk \cdot k \frac{(f_{|\vec{k}+\vec{q}|} - f_{\vec{k}})|\vec{k} - \vec{q}|}{\left| \frac{\partial E}{\partial k} \right|_{|\vec{k}+\vec{q}|} \cdot k \cdot q \cdot Sin[\theta_0]}$$

(5),

where $\theta_0$ is defined as a root of the equation $E_{\vec{k}+\vec{q}} - E_{\vec{k}} - hw = 0$ at given values of *k* and *q*. If in given point of integration mesh no real $\theta_0$ exists, the integrand of the last equation at this point is considered to be zero, because the Delta-function argument at given point will be nonzero.

From the equations for $\epsilon_{RE}$ and $\epsilon_{IM}$ (5) it follows that for the case of static screening $(w=0)$, $\epsilon_{IM}(q,w)=0$, because from the requirement of $E_{\vec{k}+\vec{q}} - E_{\vec{k}} = 0$ it follows that $f_{\vec{k}+\vec{q}} - f_{\vec{k}} = 0$.

As the dispersion law $E_k$ for HgTe quantum wells can be calculated only numerically in 8-band Kane formalism, no analytic equations for $\epsilon_{RE}(q,w)$ and $\epsilon_{IM}(q,w)$ are possible. Calculations of real and imaginary parts of dielectric function should be done numerically, based on the particular form of dispersion equation.

### III. ENERGY SPECTRA OF HGTE QUANTUM WELLS

Calculations of energy spectra are provided in the framework of the envelope functions approach when the carrier wave-function is expanded on the basis of eight Bloch band-edge (in-plane k = 0) functions [5]. The system is assumed to be periodical, although the barrier well width is chosen to correspond to the isolated quantum well.

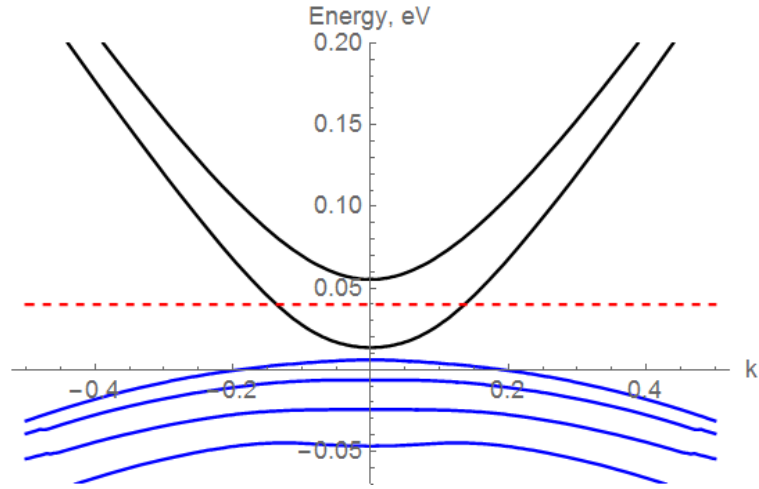

**Figure 1.** *Energy spectrum of localized carriers in $Hg_{0.15}Cd_{0.85}Te$ / $HgTe$ / $Hg_{0.15}Cd_{0.85}Te$ quantum well at T = 77 K. Dashed line represents the position of Fermi level in considered sample. The horizontal axis represents two-dimensional wave-vector k in the units of $nm^{-1}$. Black curves represent conduction band, while blue curves represent valence band.*

Calculations were made for the quantum well in (100) crystal plane, thus z-axis coincides with the growth direction of the well. Energy levels and envelope-function coefficients were found from the simultaneous solution of the system of coupled differential equations – Hamiltonian equation and boundary conditions system [5].

Calculated energy spectrum of localized carriers in $Hg_{0.15}Cd_{0.85}Te$ / $HgTe$ / $Hg_{0.15}Cd_{0.85}Te$ quantum well at temperature 77 K is shown at Fig. 1. One can see that inverted bands order is formed inside the well, because band gap between electron and hole levels is absent. Small gap between ground electron and ground holes' level is explained by the presence of mismatch strain and quantization of energy spectrum in two-dimensional medium.

Due to the proximity of electron and heavy holes bands and doping of the barriers, Fermi level is shifted high into the conduction band. The distance between Fermi level and the bottom of first electron level is greater than $2k_BT$. Also the distance from Fermi level to the top of ground heavy holes level is about $5k_BT$. Thus for this sample, we can suggest that ground level electrons only take part in the screening processes.

Electron levels in Fig. 1 are close to parabolic in the center of the zone, while they become almost linear at high values of wave-vector k. Thus, such system cannot be considered in parabolic approximation (as classical semiconductors).

There exists experimental evidence that transport properties of two-dimensional HgTe layers can be explained by graphene-like type of screening of 2DEG [2]. Thus, comparison of graphene and HgTe QW specific dielectric functions of 2DEG and discussion about their qualitative differences is one of the key points of this study.

## IV. RESULTS AND DISCUSSION

In the general case, dielectric function of 2DEG depends on scattering wave-vector $q$ and frequency $w$. However such scattering mechanisms as charged impurities scattering and electron-hole scattering are described by static screening (when frequency $w=0$).

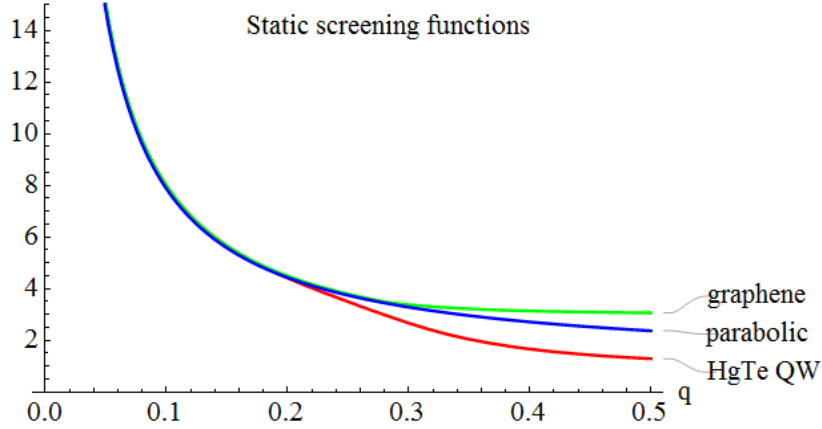

*Figure 2. The comparison of static screening functions for the graphene monolayer, QW of the thickness 20 nm with parabolic dispersion and HgTe QW of the same thickness at T = 77 K. Horizontal axis represents two-dimensional wave-vector q in the units of $nm^{-1}$.*

Three static screening functions– for QW with parabolic dispersion, for graphene and for HgTe QW, are compared in Fig. 2 for the temperature 77 K. One should note that in long-wavelength limit, when scattering wave-vector $q$ goes to zero ($q \to 0$), all three functions coincide. However, at large values of scattering wave-vector $q$, dielectric function for HgTe quantum well is several times smaller than the graphene one, which causes smaller damping of large-angle scattering on charged centers.

To make a comparison between screening processes in HgTe QW states and graphene Dirac states, all static and dynamic screening functions for graphene represented in next Figures were calculated using methods and formulas from [3], while substituting the Fermi level and actual density of states at this level from the actual HgTe QW sample.

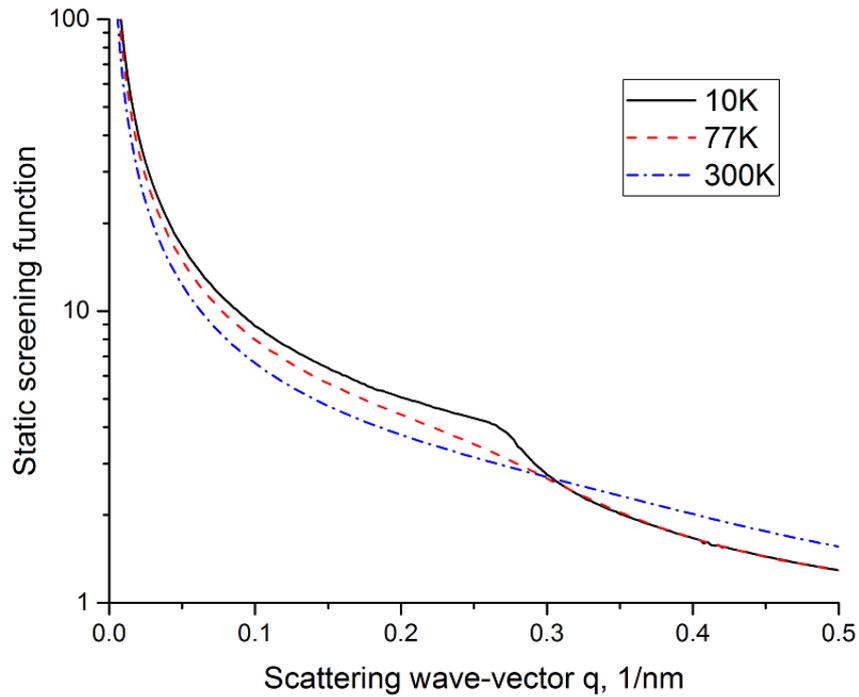

*Figure 3. Static screening functions for the HgTe QW, calculated at three different temperatures – 10 K, 77 K and 300 K. Horizontal axis represents two-dimensional wave-vector q.*

Similar dependencies of static screening function for 20 nm width HgTe quantum well for three different temperatures (10 K, 77 K, 300 K) are presented at Fig. 3. One can see that in the long-wavelength limit (when scattering wave-vector q approaches zero), screening functions for all three temperatures coincide. Growth of q leads to the decrease of the screening functions for all temperatures. It is interesting to note the appearance of bending at the curve for T = 10 K at Fig. 3. The value of Fermi wave-vector for this temperature is 0.14 nm$^{-1}$. Discussed bending is located near q = 2*$k_F$ value, while its sharpness is decreased with the temperature growth.

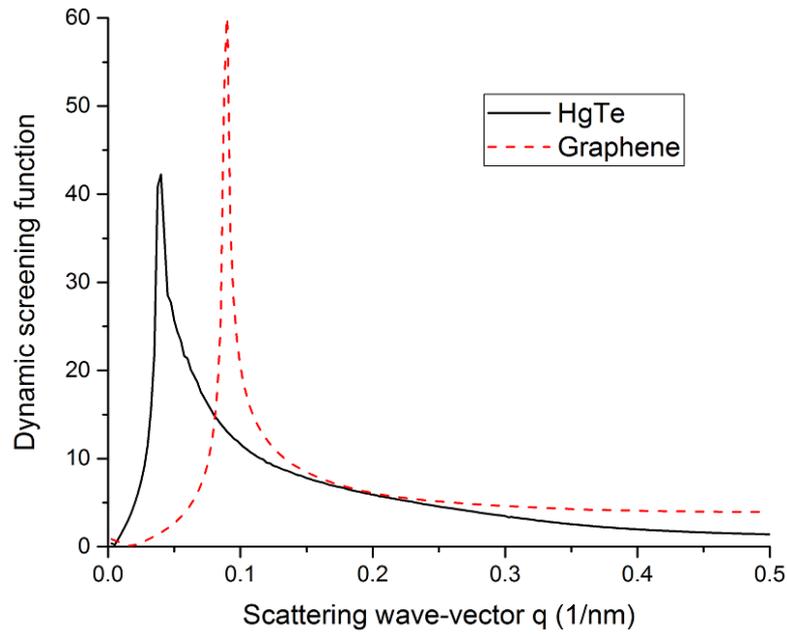

*Figure 4.* *Dynamic dielectric functions for HgTe quantum well and graphene. Calculations are provided for the frequency of longitudinal optical phonon with the energy 17 meV at T = 77 K. Horizontal axis represents two-dimensional scattering wave-vector q in the units of $nm^{-1}$.*

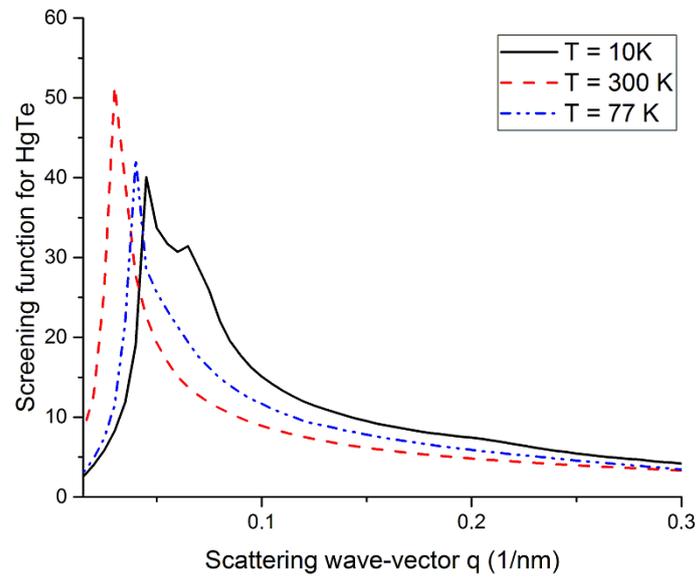

*Figure 5.* *Dynamic screening function for HgTe QW at the frequency of longitudinal optical phonon with the energy 17 meV at T = 10 K, 77 K and 300 K. Horizontal axis represents two-dimensional wave-vector q in the units of $nm^{-1}$.*

Dynamic screening functions for HgTe QW and graphene, at the frequency of optical phonon with the energy 17 meV (LO frequency), are presented in Fig. 4. It is interesting to note that transition from purely linear graphene spectrum to nonparabolic HgTe spectrum leads to the removal of singularity point from the graph and increase of the width of peak. In addition, the position of peak shifts to smaller values of $q$. Thus, HgTe QW shows stronger screening for the scattering processes of electron on optical phononsin comparison to graphene dielectric function.

Other interesting point is the dependence of dynamic HgTe QW screening function on the temperature. Dependencies of such function on scattering wave-vector at LO frequency, for three different temperatures, are plotted in Fig. 5. One can see that the increase of temperature shifts the position of the maximum to higher values of q, while decreasing its height. Nevertheless, the form of the dependence of screening function on q in general remains the same. Consequently the heating of electron, which takes place for example in THz hot-electron bolometers, should not result into sufficient changes in the 2DEG screening phenomena. The nature of furcation of the peak for T = 10 K is not clear now however the increase of calculation accuracy does not lead to its disappearance.

In the final part of the article, there we present the results for the frequency dependencies of HgTe QW and graphene screening functions.

Fig. 6 represents frequency dependencies of graphene screening function for different values of scattering wave-vector q.Each curve has sharp maximum, while small minimum emerges to the right of the peak. At high frequencies, all curves become constant and coinciding ones.

Fig. 7 represents appropriate frequency dependencies of screening functions for HgTe quantum well. Small pulses at the curves could be explained by the computational errors. Comparing the data from Fig. 7 with data from Fig. 6, one can note several important differences.

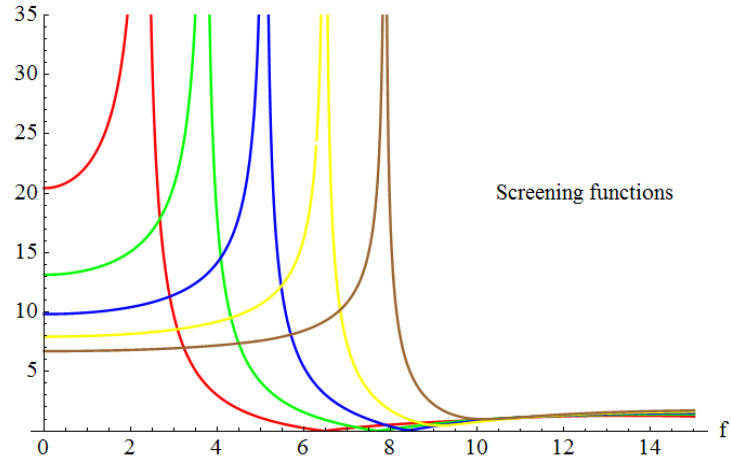

*Figure 6.* *Linear frequency dependencies of screening function for graphene. Horizontal axis represents frequency in THz. Each of the curves is buldt for the fixed value of scattering wave-vector q. From the left to the right, appropriate scattering wave-vectors of these curves are 0.05, 0.08, 0.11, 0.14 and 0.17 nm$^{-1}$.*

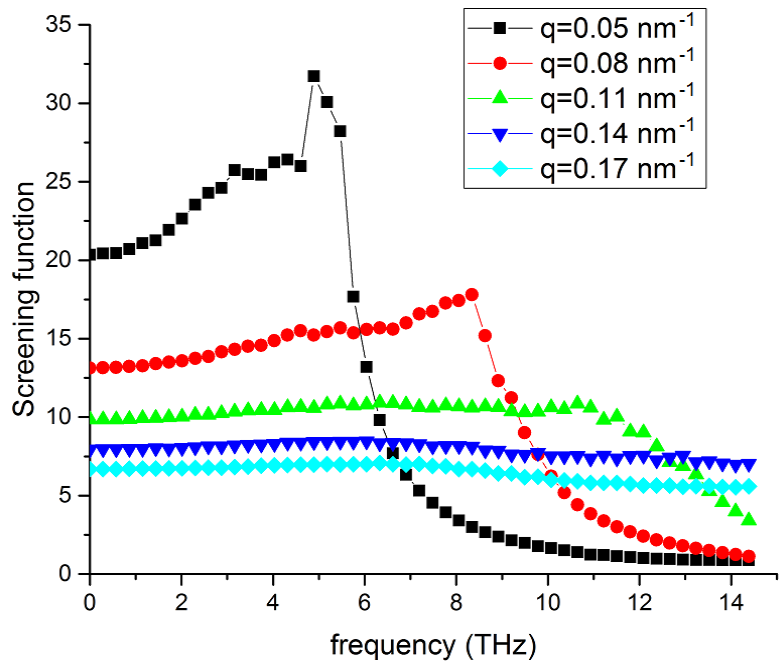

*Figure 7.* *Linear frequency dependencies of screening function for HgTe quantum well. Horizontal axis represents frequency in THz. Each of the curves is built for the fixed value of scattering wave-vector q. From the higher to the lower curves, appropriate scattering wave-vectors of these curves are 0.05, 0.08, 0.11, 0.14 and 0.17 of nm$^{-1}$.*

Transition from purely linear (graphene) spectrum to nonparabolic (HgTe QW) spectrum leads to the removal of narrow peaks of infinite height. They are replaced by wide plateaus, which height is decreased during the increase of scattering wave-vector $q$, while width of these plateaus is sufficiently broader than the width of corresponding peaks at Fig. 4. As one can see on two highest curves (which correspond to the lowest values of $q$), maximum value frequencies of HgTe QW screening function are bigger than corresponding frequencies from Fig. 6.

From the comparison of curves from Fig. 6 and Fig. 7 corresponding to the same value of scattering wave-vector, we can make several conclusions.

For the frequencies that are larger than peak value frequency, 2DEG screening in HgTe QW is much more efficient, than in graphene case. On the other hand, no frequencies do exist in HgTe QW, for which absolute screening (e.g. infinitely large screening function) of 2DEG is possible.

## V. PLASMONS IN RPA

### A. *Problem overview*

Two-dimensional plasmons or collective oscillations of the density of 2DEG provide the mechanism to absorb incoming radiation at the frequencies order of 1 THz. This feature makes them perspective for design of THz detectors. As inverted bands HgTe quantum wells can provide high electron mobility, low channel resistance and noise, calculation of their plasma frequencies is quite important.

In the literature, they usually model the situation when only interface states are utilized for the carrier transport [6]. In the case of topological insulator, the interface levels of QW lie in the band gap of the energy spectrum of "bulk" 2D carriers. Such interface levels are characterized by Dirac-like energy dispersion originating from the sign reversal of carrier mass at the interface. However, such interface plasmons are essential only when transport via bulk levels of QW is suppressed, which means that Fermi level is in the middle of band gap ofQW, while such band gap is much greater than $k_BT$. This results into the usage of very low temperatures [7] or heavily strained HgTe QW samples with wide band gap (up to 200 meV) [8]. Both approaches are undesirable for construction of real devices, because deep cooling results into bulky and expensive devices, while heavily strained samples could be affected by fast degradation.

Also in the literature there exist numerical calculations describing plasmons in the bulk levels of HgTe quantum well [9, 10]. However they are made in the parabolic approximation, neglecting bands hybridization and non-parabolicity of energy spectrum of bulk carriers. Moreover, [9] is done for $T = 0$. In [10] temperature is not mentioned in text or in formulas, thus one could suggest that this work is also done for $T = 0$.

Thus up to our knowledge, no numerical simulations of frequency dispersion of "bulk" plasmons (existing at bulk levels of HgTe QW) can be found at the literature for $T = 77$ K.

Nevertheless, there exist experiments which allow one to suggest that such plasmons in HgTe QW do exist at liquid nitrogen and higher temperatures [11].

There are presented our numerical results for the frequency dispersion of "bulk" plasmons in HgTe QW at 77 K in the next subsection. These results properly incorporate bands mixing and non-parabolicity of dispersion law.

B. *Our numerical results for plasmons in RPA*

Calculation of screening function in the RPA framework has the immediate theoretical consequence for plasmonic frequencies. Due to the physics incorporated in the RPA, zero values of dielectric screening function (1) correspond to the collective excitations of 2DEG. Thus the calculation of frequencies and scattering wave-vectors, at which screening function is zero, gives us the plasmon dispersion spectrum of the QW. We will carry such calculations in long-wavelength limit ($q \to 0$).

In this limit, real part of the screening function (5) can be rewritten as:

$$\epsilon_{RE}^{long-wavelength}(\vec{q},w) = 1 + \frac{e^2}{\varepsilon_0 q\pi} \iint \frac{dk \cdot d\theta \cdot k \cdot q \cdot Cos(\theta)\frac{df_k}{dk}}{\hbar w - q \cdot Cos(\theta) \cdot \frac{dE_k}{dk}} \qquad (6)$$

Imaginary part of the screening function (5) is calculated via one-dimensional integral, thus it does not need any further simplification, as its integration error is already much smaller than for the real part.

Plasmon frequencies for the HgTe QW and single-layer graphene are presented in Fig. 8. Both calculations are carried out in the long-wavelength limit. Calculations for HgTe QW are provided via (6), (5) at T = 77 K. Calculations for graphene are provided using formula (18) from [3] at T = 0 K. One can see that both dependencies are presented as parallel straight lines in double-logarithmic scale.

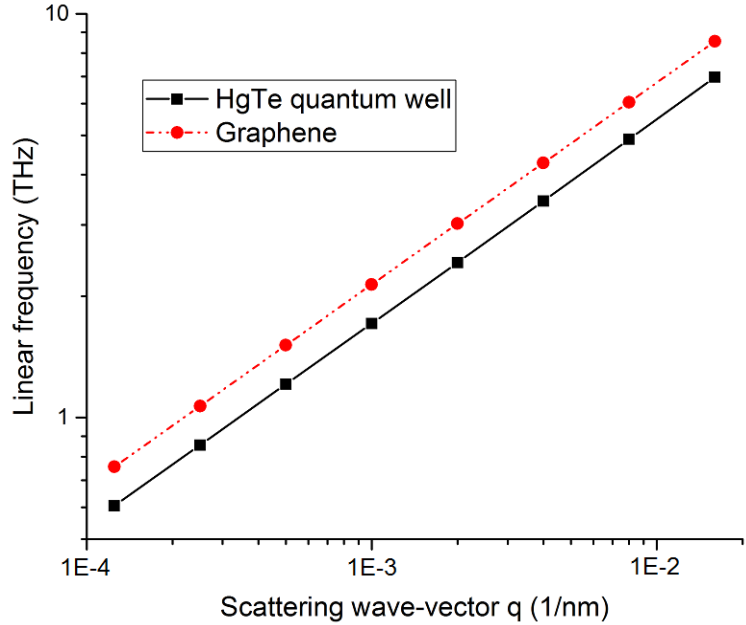

*Figure 8. Plasmon frequencies (in THz) for HgTe QW and graphene, in the dependence on scattering wave-vector q. Plasmon frequencies for HgTe QW are calculated at 77 K. Plasmon frequencies for graphene are calculated at T = 0 K using formula (18) from [3].*

The plasmons following from our RPA calculations are not interface plasmons, because they are based of the 2D wave-function of electrons localized inside the well. As our RPA calculations are carried out for temperatures much greater than liquid helium temperature, appropriate plasmons are referred in this work as "high-temperature" plasmons.

Another important difference of high-temperature plasmons of this work and low-temperature interface plasmons from the literature lies in the coupling of electron spin and electron motion direction. In this case, near the phase transition point four-band Dirac Hamiltonian describes the energy spectrum of HgTe QW at the liquid helium temperature [13]. Such description allows one to present Hamiltonian in the form when its right-lover part is the complex conjugate of its left-upper part, while left-lower and right-upper parts of Hamiltonian being zero. From such form of Hamiltonian, one can note the presence of time-reversal symmetry of the system. It means that the direction of electron motion and the direction of its spin are coupled (in the absence of magnetic field and magnetic impurities), plasma waves are accompanied by spin waves, while the total phenomena is called spin-plasmons.

In the contrast, even at the liquid nitrogen temperature and far from the phase transition point, the energy spectrum cannot be modeled in 4-band approximation. 8-band k.p "high-temperature" plasmons calculations of the energy spectrum do not allow presenting Hamiltonian in the "time-reversal" form. This means that in the general case, direction of electron motion and direction of electron spin are not coupled, while corresponding plasmons are not spin-plasmons.

## VI. CONCLUSIONS

We set ourselves the task of studying the behavior of the dielectric screening function of bulk type electrons and plasmons in the inverted bands HgTe quantum wells at high temperatures and provide the comparison of the screening function for the Schrodinger fermions in this structure with the appropriate screening function for graphene monolayer with the Dirac fermions. The Random-Phase Approximation (RPA) dielectric screening function for the inverted bands HgTe QW, which takes into account realistic 8-bands k.p Hamiltonian for energy spectrum, was numerically simulated in this work.

It was found that for the case of static screening, that have place for the electron scattering on charged impurities and holes, our dielectric function is similar to graphene one in the long-wavelength limit. However, at the large values of scattering wave-vector q, the dielectric function for HgTe quantum well is several times smaller than the graphene one, which causes to the weaker damping of large-angle scattering on charged centers, and, as a result, leads to the smaller mobility.

In the case of dynamical screening, that describes the process of scattering on longitudinal optical phonons, there exists a maximum on the dependence of dielectric function on scattering vector amplitude $q$, both for graphene and HgTe dielectric functions, for the fixed frequency. For the graphene case this maximum growth to infinity. However, for HgTe dielectric function, this maximum is shifted to sufficiently smaller values of $q$, while its height is finite, which affects the scattering on longitudinal optical phonons in the well.

Considering the dependence of screening function on the frequency at fixed values of wave-vector $q$, we can outline several important differences between graphene and HgTe QW screening functions. For graphene, there exists sharp maximum on the graph; also there exists minimum which is smoothed with increasing of $q$. In the case of HgTe QW such maximum of the screening function is strongly dumped, forming the plateau. The peak value of this plateau is smaller than peak value at graphene graph; however width of plateau is substantially broader than the width of peak at graphene graph.

We also simulated the plasmons' frequencies of 2DEG in inverted HgTe QW in the long-wavelength limit for T = 77 K. The plasmons in our case of HgTe QW are not the interface plasmons, because they are based on the 2D electron wave-function localized inside the well. As our RPA calculations are carried out for temperatures much greater than liquid helium temperature, appropriate plasmons are referred in this work as "high-temperature" plasmons. Plasmon frequencies on scattering vector dependencies are represented as parallel straight lines in double-logarithmic scale for the HgTe QW and single-layer graphene.

Thus, all these peculiarities need to be taken into account when determining the response of these structures to external high frequency radiation when creating detector structures.

*Acknowledgment*

This research was partly supported by VW Stiftung Project „Opto-electronic and transport phenomena in narrow gap semiconductor structures for terahertz detection" and NAS of Ukraine Program NANO №11/17-H.

*References*

[1] E. O. Melezhik, J. V. Gumenjuk-Sichevska, F. F. Sizov, "Modeling of Noise and Resistance of Semimetal $Hg_{1-x}Cd_xTe$ Quantum Well used as a Channel for THz Hot-Electron Bolometer", *Nanoscale Research Letters* 11(1), p. 181 (2016).

[2] C. Brüne, C. Thienel, M. Stuiber, J. Böttcher, H. Buhmann, E. G. Novik, Chao-Xing Liu, E. M. Hankiewicz, and L. W. Molenkamp, "Dirac-Screening Stabilized Surface-State Transport in a Topological Insulator", *Phys. Rev. X* 4, p. 041045 (2014).

[3] E. H. Hwang, S. Das Sarma, "Dielectric function, screening, and plasmons in two-dimensional graphene", *Physical Review B* 75(20), id. 205418 (2007).

[4] T. Ando, A.B. Fowler and F. Stern, "Electronic properties of two-dimensional systems", *Rev. Mod. Phys.* 54, p. 437 (1982).

[5] E. O. Melezhik, J. V. Gumenjuk-Sichevska and F. F. Sizov, "Modeling of electron energy spectra and mobilities in semi-metallic $Hg_{1-x}Cd_xTe$ quantum wells", *Journal of Applied Physics* 118(19), p. 194305 (2015).

[6] Yi-Ping Lai, I-Tan Lin, Kuang-Hsiung Wu and Jia-Ming Liu, "Plasmonics in Topological Insulators", *Nanomaterials and Nanotechnology* 4(13), doi: 10.5772/58558 (2014).

[7] Clement Bouvier, Tristan Meunier, Roman Kramer, Laurent P. Levy, Xavier Baudry and Philippe Ballet, "Strained HgTe: a textbook 3D topological insulator", arXiv, https://arxiv.org/abs/1112.2092v1 (2011).

[8] Jin Li, Chaoyu He, Lijun Meng, Huaping Xiao, Chao Tang, Xiaolin Wei, Jinwoong Kim, Nicholas Kioussis, G. Malcolm Stocks & Jianxin Zhong, "Two-dimensional topological insulators with tunable band gaps: Single-layer HgTe and HgSe", *Scientific Reports* 5:14115, DOI: 10.1038/srep14115 (2015).

[9] Danhong Huang, Zhitang Lin, Shixun Zhou, "Dielectric response of a semi-infinite HgTe/CtiTe superlattice from its bulk anti surface plasmons", *Phys. Rev. B* 40(3), p. 1672 (1989).

[10] Dan-hong Huang, Jian-ping Peng and Shi-xun Zhou, "Intrasubband and intersubband plasmons in a semi-infinite Fibonacci HgTe/CdTe superlattice", *Phys. Rev. B* 40(11), p. 7754 (1989).

[11] M. L. Bansal, A. P. Roy, and Alka Ingale, "Raman scattering from coupled plasmon-phonon modes in HgTe", *Phys. Rev. B* 42(2), p. 1234 (1990).

[12] Stefan Juergens, Paolo Michetti, and Björn Trauzettel, "Screening properties and plasmons of Hg(Cd)Te quantum wells", *Phys. Rev. B* 90, p. 115425 (2014).

[13] B. A. Bernevig, T. L. Hughes, Shou-Cheng Zhang, "Quantum Spin Hall Effect and Topological Phase Transition in HgTe Quantum Wells", *Science* 314, pp. 1757-1761 (2006).

[14] J. R. Meyer, D. J. Arnold, C. A. HofFman, F. J. Bartoli, "Electron and hole in-plane mobilities in HgTe-CdTe superlattices", *Phys. Rev. B.* 46, p. 4139 (1992);

[15] A. Nafidi, "Correlation Between Band Structure and Magneto-Transport Properties in n-type HgTe/CdTe Two-Dimensional Nanostructure Superlattice. Application to Far-Infrared Detection", Chap. 6 in *Optoelectronics - Advanced Materials and Devices*, Eds. S.L. Pyshkin and J.M. Ballato, p. 145, InTech (2013).